\documentclass[prd,twocolumn,aps,psfig,nofootinbib,nobibnotes,superscriptaddress,preprintnumbers,times]{revtex4-2}
\usepackage{graphicx,bm,color,amsmath,amssymb, mathtools,subcaption}
\usepackage[dvipsnames]{xcolor}
\usepackage[hidelinks]{hyperref}
\hypersetup{
  colorlinks   = true, 
  urlcolor     = [RGB]{46,48,146}, 
  linkcolor    = [RGB]{46,48,146}, 
  citecolor    = [RGB]{46,48,146} 
}

\newcommand{\EQ}{\begin{equation}}
\newcommand{\EN}{\end{equation}}
\newcommand{\EQA}{\begin{eqnarray}}
\newcommand{\ENA}{\end{eqnarray}}
\newcommand{\eq}[1]{(\ref{#1})}

{}
{}
{}

\newcommand{\GeV}{\,{\rm GeV}}

\newcommand{\s}{\,{\rm s}}

\newcommand{\km}{\,{\rm km}}

\newcommand{\Mpc}{\,{\rm Mpc}}

\makeatletter
\renewcommand\@makecaption[2]{%
  \par
  \vskip\abovecaptionskip
  \begingroup
   \small\rmfamily
    \begingroup
     \samepage
     \flushing
     \let\footnote\@footnotemark@gobble
     \@make@capt@title{#1}{#2}\par
    \endgroup
  \endgroup
  \vskip\belowcaptionskip
}
\makeatother

\usepackage{braket,orcidlink}

\begin{document}

\title{Stochastic gravitational wave background detection using NANOGrav 15-year data set in the context of massive gravity}
\date{\today}
\author{Chris~Choi\,\orcidlink{0009-0005-2328-3044}}
\email{Contact author: minyeonc@andrew.cmu.edu}
\affiliation{McWilliams Center for Cosmology and Astrophysics and Department of Physics, \href{https://ror.org/05x2bcf33}{Carnegie Mellon University}, Pittsburgh, Pennsylvania 15213, USA}

\author{Jacob~Magallanes\,\orcidlink{0009-0004-0616-146X}}
\email{Contact author: jmagalla@andrew.cmu.edu}
\affiliation{McWilliams Center for Cosmology and Astrophysics and Department of Physics, \href{https://ror.org/05x2bcf33}{Carnegie Mellon University}, Pittsburgh, Pennsylvania 15213, USA}

\author{Murman~Gurgenidze\,\orcidlink{0009-0008-7197-2890}}
\email{Contact author: mgurgeni@andrew.cmu.edu}
\affiliation{McWilliams Center for Cosmology and Astrophysics and Department of Physics, \href{https://ror.org/05x2bcf33}{Carnegie Mellon University}, Pittsburgh, Pennsylvania 15213, USA}
\affiliation{School of Natural Sciences and Medicine, \href{https://ror.org/051qn8h41}{Ilia State University}, 0194 Tbilisi, Georgia}

\author{Tina~Kahniashvili\,\orcidlink{0000-0003-0217-9852}}
\email{Contact author: tinatin@andrew.cmu.edu}
\affiliation{McWilliams Center for Cosmology and Astrophysics and Department of Physics, \href{https://ror.org/05x2bcf33}{Carnegie Mellon University}, Pittsburgh, Pennsylvania 15213, USA}
\affiliation{School of Natural Sciences and Medicine, \href{https://ror.org/051qn8h41}{Ilia State University}, 0194 Tbilisi, Georgia}
\affiliation{\href{https://ror.org/02gkgrd84}{Abastumani Astrophysical Observatory}, Tbilisi GE-0179, Georgia}

\begin{abstract}
Convincing evidence of a stochastic gravitational wave (GW) background has been found by the NANOGrav Collaboration in the 15-year data set. From this signal, we can evaluate the possibility of its source being from the early Universe through the tensor perturbations induced by a massive spin-2 graviton field. We consider a time-dependent model of the minimal theory of massive gravity and find values of the graviton mass, mass cutoff time, and Hubble rate of inflation that amplify the energy spectra of primordial GWs sufficiently to reproduce the signal from the NANOGrav data within 1-3 standard deviation. However, a suppression mechanism for high-frequency modes must be introduced to conservatively obey the big bang nucleosynthesis (BBN) bound. While there are regions of the parameter space that reproduce the signal, it remains a challenge to simultaneously respect the BBN and cosmic microwave background bounds without making the graviton mass cutoff time too deep into the matter-dominated era.
\end{abstract}

\maketitle
\section{Introduction}
Evidence supporting the existence of a stochastic gravitational wave background (SGWB) has been found in 15 years of observation of pulsars by the North American Nanohertz Observatory for GWs (NANOGrav Collaboration) \cite{Agazie:2023}, as well as by the Chinese Pulsar Timing Array (CPTA) \cite{Xu:2023wog}, the European Pulsar Timing Array (EPTA) \cite{EPTA:2023sfo,EPTA:2023akd,EPTA:2023fyk}, and the Parkes Pulsar Timing Array (PPTA) \cite{Zic:2023gta,Reardon:2023gzh}. We investigate alternatives to the astrophysical explanation (like inspiraling supermassive black hole binaries (SMBHBs) \cite{Rajagopal:1995,Jaffe:2002rt,Burke-Spolaor:2018bvk}), such as the time-dependent model of massive gravity (MG) proposed in Ref.\ \cite{Fujita:2018ehq}. 

Compared to astrophysical origins, more exotic explanations lie in cosmological sources of the SGWB  \cite{Maggiore:1999vm, Caprini:2018mtu, Chen:2021wdo, Wu:2021kmd, Chen:2021ncc, PPTA:2022eul, Wu:2023pbt, Wu:2023dnp, Madge:2023cak, Ellis:2023oxs, Bian:2023dnv, Figueroa:2023zhu}. Such cosmological explanations for the source include  cosmic strings \cite{Damour:2004kw,Siemens:2006yp, Chen:2022azo,Bian:2022tju, Antusch:2024nqg}, domain walls \cite{Ferreira:2022zzo, Zhang:2023nrs, Gouttenoire:2023ftk}, first-order phase transitions in the early Universe \cite{Kibble:1976sj, Vilenkin:1984ib,Caprini:2010xv, Kobakhidze:2017mru, Arunasalam:2017ajm, Xue:2021gyq, NANOGrav:2021flc, Moore:2021ibq, Addazi:2023jvg, Athron:2023xlk, Bringmann:2023opz, Ashoorioon:2022raz, Gouttenoire:2023bqy}, primordial magnetic fields \cite{Neronov:2020qrl,Brandenburg:2021tmp,RoperPol:2022iel,Kahniashvili:2021gym}, primordial GWs \cite{Grishchuk:1976, Grishchuk:1977zz, Starobinsky:1980te, Linde:1981mu, Fabbri:1983us, Grishchuk:2005qe, Lasky:2015lej, Kawai:2023nqs, Basilakos:2023xof, Basilakos:2023jvp, Benetti:2021uea, Vagnozzi:2020gtf}, scalar-induced GWs \cite{Tomita:1967non, Saito:2008jc, Young:2014ana, Yuan:2019udt, Yuan:2019wwo, Chen:2019xse, Cai:2019bmk, Yuan:2019fwv, Liu:2021jnw, Liu:2023ymk, Cai:2023dls, Choudhury:2023fjs, Choudhury:2023fwk, Bhattacharya:2023ysp, Choudhury:2023hfm, Kawai:2021edk, Pi:2021dft, Domenech:2020ers, Wang:2023sij} generated by primordial black holes \cite{Zeldovich:1967lct,Hawking:1971ei,Carr:1974nx,Chen:2018czv,Chen:2018rzo,Liu:2018ess,Liu:2019rnx,Chen:2019irf,Liu:2020cds,Wu:2020drm,Chen:2021nxo,Chen:2022fda,Chen:2022qvg,Liu:2022iuf,Zheng:2022wqo, Choudhury:2013woa, Franciolini:2023pbf, Wang:2023ost}, and even non-GW explanations \cite{Chowdhury:2023xvy}. The NANOGrav and EPTA Collaborations have considered some of these aforementioned new physics sources \cite{Afzal:2023,EPTA:2023xxk}. In this paper, we consider the primordial GW hypothesis, with an amplification due to MG. We consider MG as a promising alternative to explain the present exponential expansion.

Primordial GWs generated during inflation, being generated through parametric resonance amplification of quantum mechanical fluctuation, and freely propagating in the radiation-dominated plasma, are predicted by theory 
\cite{Grishchuk:1976,Grishchuk:1977zz,Starobinsky:1980te,Linde:1981mu,Fabbri:1983us}. We can expect that the signatures of these waves we detect differ from what we expect based on general relativity (GR). In fact, the generation and propagation of these primordial GWs could very well be modified by alternate theories of gravity, including MG, first introduced by Fierz and Pauli in 1939 \cite{Fierz:1939ix}. Since then, many attempts have been made to construct a consistent nonlinear theory of MG. Any purely linear theory suffers from the van Dam-Veltman-Zakharov discontinuity \cite{vanDam:1970vg,Zakharov:1970cc}, which prevents the theory from reducing to GR in the massless limit. Attempts have been made to address this, like the nonlinear extensions to the Fierz-Pauli theory that exhibit the Vainshtein mechanism \cite{Vainshtein:1972sx}. This has its own set of problems, like the Boulware-Deser ghost and other ghost degrees of freedom \cite{Boulware:1972yco,Dubovsky:2004sg}. This presents a significant obstacle for trying to come up with a consistent theory.

Recently, attempts have been made to provide ghost-free nonlinear MG  theories, such as the de Rham-Gabadadze-Tolley (dRGT) theory \cite{Hassan:2011tf, Hassan:2011ea, deRham:2010ik,deRham:2010kj}. One of the consequences of the dRGT model is that there are no isotropic solutions at large cosmological scales. Since the isotropy of the Universe is a fundamental symmetry, several attempts have been made to avoid anisotropic solutions and retain the stability of perturbations. These attempts include quasidilaton MG \cite{DAmico:2012hia}, where it was shown to have unstable perturbations around a self-accelerating background \cite{Gumrukcuoglu:2013nza}. A further extension, where one also allows for a new type of coupling between the massive graviton and the quasidilaton \cite{DeFelice:2013tsa}, admits a stable self-accelerated solution. Despite progress in building quasidilaton MG models, it appears that the Boulware-Deser ghost still remains present \cite{Mukohyama:2013raa}. Other extensions of the dRGT model include the minimal theory of MG (MTMG) \cite{DeFelice:2015hla,DeFelice:2015moy}, the Dirac-Born-Infeld-dRGT \cite{Kazempour:2022giy}, and the Brans-Dicke-de Rham-Gabadadze model \cite{Kazempour:2022let}.

MTMG proposes a novel modification of dRGT: doing away with Lorentz invariance and allowing gravitons to form a 3D rotation group, e.g., \cite{Arkani-Hamed:2003pdi,Rubakov:2004eb,Dubovsky:2004sg,Blas:2009my,Rubakov:2008nh,Blas:2007zz,Comelli:2013txa,Langlois:2014jba}. This no longer requires the existence of five degrees of freedom, and due to the lack of observations of new propagating degrees of freedom, we consider MTMG in this paper. In this model, we no longer have to consider the Higuchi bound, which ordinarily requires that the ratio of the mass of the graviton to the Hubble scale of inflation be up to an order of unity \cite{Higuchi:1986py}. 
This model allows for stable, isotropic Friedmann-Lema\^{\i}tre-Robertson-Walker (FLRW) cosmologies, which other models suffer instabilities in, thereby leading to an experimentally viable cosmology \cite{DeFelice:2015hla}.

In this paper, we consider a time-dependent MTMG, specifically one in which the graviton mass is a step function of time as described in Ref.\ \cite{Fujita:2018ehq} (hereafter the step function mass (SFM) model). We calculate the energy density of primordial GWs created during inflation, at the present time, in the presence of MG and compare it to the signals we observed in the 15-year NANOGrav data set (hereafter NG15). The goal of this paper is to show whether MG, through primordial GWs, is able to reproduce the observed SGWB.

The paper is arranged as follows: in Sec.\ \ref{sec:setup}, we introduce the model and the assumptions we make. In Sec.\ \ref{sec:energy}, we derive the energy density and discuss its behavior. In Sec.\ \ref{sec:results}, we compare the model to the signals detected by NANOGrav. In Sec.\ \ref{sec:discussion}, we discuss the implications of our findings and discuss future work. Throughout this paper, we use  (--,+,+,+) for the Minkowski metric $\eta_{\mu\nu}$, and we use natural units and set $c = $$\ \hbar = $$\ k_B = $$\ 1$. We also set the present-day Hubble parameter $H_0 = h_0\ 100 \km \s^{-1}\Mpc^{-1} = 67.66 \km \s^{-1}\Mpc^{-1}$ and the present-day density parameters for radiation, matter, curvature, and dark energy $\{\Omega_r, \Omega_m, \Omega_k, \Omega_\Lambda\} = \{9.182\times10^{-5},0.3111,0,0.6889\}$ to match the latest \textit{Planck} 2018 TT, TE, EE + lowE + lensing + BAO data \cite{Planck:2018vyg}.

\section{Model Setup}\label{sec:setup}
We start with defining the FLRW metric $g_{\mu\nu}$,
\begin{equation}\label{eqn:metric}
    g_{\mu\nu}dx^{\mu} dx^{\nu} = -N^2(t)dt^2 +a^2(t)\left(\frac{dr}{1-Kr^2} + r^2 d\Omega^2\right),
\end{equation}
where $t$ is proper time, $K$ is the spatial curvature, $d\Omega^2 = d\theta^2 + \sin^2\theta d\phi^2$ where $\theta$ and $\phi$ are the polar and azimuthal angles in spherical coordinates, $N(t)$ is the lapse, and $a(t)$ is the scale factor. We then consider the general Lorentz-violating action for MTMG [Eq.\ (61) of Ref.\ \cite{DeFelice:2015moy} ],
\begin{equation}\label{eqn:action}
    S = S_{\text{pre}} + S_{\text{MT}} + S_{\text{mat}},
\end{equation}

\hspace{-1em}where $S_{\text{pre}}$ is the precursor action that MTMG is constructed from (equivalent to the dRGT action with vielbeins in the ADM formalism \cite{Arnowitt:1962hi} substituted into Eq.\ (11) of Ref.\ \cite{DeFelice:2015moy}), $S_{\text{MT}}$ is the additional action in MTMG, and $S_{\text{mat}}$
is the action for the matter fields. $S_{\text{pre}}$ is defined as [Eq.\ (65) of Ref.\ \cite{DeFelice:2015moy} ],
\begin{equation} \label{eqn:action_pre}
    S_{\text{pre}} = S_{\text{GR}} + \frac{M_{\text{Pl}}^2}{2}\sum_{i=1}^4 d^4x \mathcal{S}_i,
\end{equation} 
where $S_{\text{GR}}$ is the action for GR, $M_{\text{Pl}}$ is the Planck mass, and $\mathcal{S}$ is defined for the graviton mass term in Eqs.\ (66)-(69) in Ref.\ \cite{DeFelice:2015moy}. We find that the definition of $S_\text{pre}$ is sufficient for this paper, since we only consider the general quadratic action of MTMG, which happens to be the same as dRGT's. Our quadratic action is thus [Eq.\ (13) of Ref.\ \cite{DeFelice:2015hla} ],
\begin{equation} \label{eqn:action_quad}
     S = \frac{M_{\text{Pl}}^2}{8}\int d^4xNa^3\left[ \frac{h'^2_{ij}}{N^2}- \frac{(\partial h_{ij})^2}{a^2} - M_{\text{GW}}^2h_{ij}^2 \right],
\end{equation}
where $h_{\mu\nu} = h_{\mu\nu}(t,{\bf x})$ is the metric perturbation defined by $g_{\mu\nu} = \eta_{\mu\nu} + h_{\mu\nu}$, the primes ($\,'$) indicate derivatives with respect to $t$, and we have defined $M_\text{GW}$ as the mass of the graviton. We decompose the spatial component of the tensor perturbation $h_{ij}$ into its helicity states [Eq.\ (19.214) of Ref.\ \cite{Maggiore:v2} ], 
\begin{equation}\label{eqn:decomp}
    h_{ij}(\tau, {\bf k}) = \sum_{\lambda \in \{+, \times\}}e^\lambda_{ij}(\hat{\bf k})h_k^\lambda(\tau, {\bf k}),
\end{equation}
where ${\bf k}$ is the comoving momentum, $\tau$ is the conformal time defined by $\tau = \int \frac{N(t)}{a(t)}dt$, and $e^\lambda_{ij}$ are the polarization tensors defined by 
\begin{equation}\label{eqn:polarization}
    e^+_{ij}(\hat{\bf k}) = \hat{\bf u}_i\hat{\bf u}_j - \hat{\bf v}_i\hat{\bf v}_j, \quad e^\times_{ij}(\hat{\bf k}) = \hat{\bf u}_i\hat{\bf v}_j - \hat{\bf v}_i\hat{\bf u}_j,
\end{equation}
given in Eqs.\ (1.54)-(1.56) of Ref.\ \cite{Maggiore:v1}. Here, $\hat{\bf u}, \hat{\bf v}$ are the unit vectors orthogonal to the direction of propagation $\hat{\bf k}$ and to each other. We briefly note that we have only two propagating modes of freedom in MTMG, allowing stable FLRW cosmologies, whereas dRGT produces ghosts in the helicity-0 and helicity-1 sectors, thus unstable FLRW cosmologies \cite{DeFelice:2015hla}. After we minimize the action from Eq.\ (\ref{eqn:action_quad}) and take into account the helicity decomposition in Eq.\ (\ref{eqn:decomp}), we obtain the equation of motion for $h_k^\lambda$ (with the helicity modes suppressed since they have the same equation of motion)
\begin{equation}\label{eqn:eom}
    \bar{h}_k'' + \left(c_g^2(\tau) k^2 + a^2 M_\text{GW}^2 - \frac{a''}{a} + 2Kc_g^2(\tau)\right)\bar{h}_k = 0,
\end{equation}
where $\bar{h}_k = ah_k$, the primes ($\,'$) have been redefined to indicate derivatives with respect to $\tau$, and $c_g(\tau)$ is the effective sound speed associated with GWs and may be dependent on time. For this paper, we set $K = 0$ and $c_g = 1$.\footnote{Refer to Sec.\ B2 of Ref.\ \cite{Gumrukcuoglu:2012wt} for consideration of nonzero $K$ and Sec.\ IV E4 of Ref.\ \cite{Gumrukcuoglu:2012wt} for consideration of general $c_g$.}

\begin{figure}[t]
\centering
\includegraphics[scale=0.75]{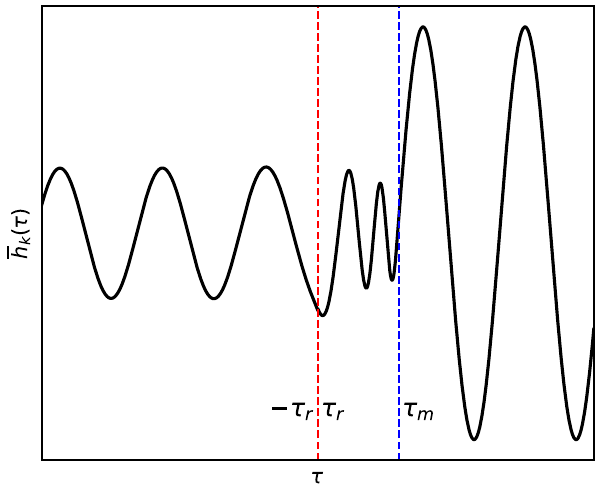}
\caption{Evolution of the real part of $\bar{h}_k(\tau)$. Note that conformal time jumps from $-\tau_r$ to $\tau_r$. This is a numerical solution to Eq.\ (\ref{eqn:eom}), and the exact values for the parameters are detailed in the source code \cite{Choi:2024}.}
 \label{fig:mode}
\end{figure}

We consider graviton masses on the order of the Hubble scale, and so the evolution of the GWs during inflation will be important.
The scale factor, as described in Eq.\ (4) of Ref.\ \cite{Fujita:2018ehq}, is
\begin{equation}\label{eqn:scale_fac}
    a(\tau) = 
    \begin{cases}
        -1/(H_{\inf}\tau) & \tau < -\tau_r \\
        a_r \tau/\tau_r & \tau > \tau_r \\
   \end{cases},
\end{equation}
where $H_{\inf}$ is the Hubble parameter during inflation when the scale corresponding to the CMB exits the Hubble horizon, and $a_r$ is the scale factor at the reheating time $\tau_r = 1/(a_r H_{\inf})$. We assume $a_r$ is fixed in this discussion, while $H_{\inf}$ and $\tau_r$ may vary. The bounds on $H_{\inf}$ are discussed in Ref.\ \cite{Jiang:2015qor} and are respected in this paper. The conformal time jumps in value from $-\tau_r$ to $\tau_r$. This is to allow for the scale factor and its first $\tau$ derivative to remain continuous. Figure \ref{fig:mode} illustrates what a generic mode function looks like. The behavior in the three regions of $\tau$ is discussed in detail in Ref.\ \cite{Fujita:2018ehq}.

We model the graviton mass as the function given in Eq.\ (5) of Ref.\ \cite{Fujita:2018ehq},
\begin{equation}\label{eqn:mass_case}
    M_\text{GW}(\tau) = 
    \begin{cases}
        m & \tau < \tau_m \\
        0 & \tau > \tau_m
   \end{cases},
\end{equation} 
where $\tau_m$ is the conformal time when the graviton mass instantaneously drops to 0. A smoothly decaying function would be preferred, but the exact form of this function is highly model dependent and can depend on the dynamics of other fields, for example, some scalar field $\mu(\phi(\tau))$. The general phenomenological aspects of a theory of MG with a time-dependent graviton mass can still be explored with the simplest scheme for time dependence, the step function, with the benefit of simplifying numerical calculations and avoiding the intricacies related to the model-dependent scalar fields. Thus, we choose to study the SFM model.

\section{Energy Density of gravitational waves}\label{sec:energy}
When we study models of gravity, we often have to determine how GWs from the primordial era are influenced by deviations from GR. The energy density spectrum of these GWs at the present time as a function of frequency is perhaps the most practical way to observe the effects of beyond-GR theories. We want to look at the energy fraction of the GWs per logarithmic interval of $k$ at the current time. We have the following definition for the energy density:
\begin{equation}\label{eqn:omega_sfm}
    \Omega_\text{GW} = \frac{1}{\rho_c}\frac{d \rho_\text{GW}} {d \log{k}},
\end{equation}
where $\rho_c = 3H^2/8\pi G$ is the critical density, and we note that the derivative with respect to $\log k$ is a notational way of representing the spectral density of $\rho_\text{GW}$ (see footnote 65 of Ref.\ \cite{Maggiore:v1}). Our energy density today is defined in terms of the primordial tensor power spectrum $\mathcal{P}_0(f) := \mathcal{P}(\tau_0, f)$ as detailed in Eq.\ (19.288) of Ref.\ \cite{Maggiore:v2},
\begin{equation}\label{eqn:omega_0_sfm}
    \Omega_{\text{GW},0}(f) = \frac{\pi^2}{3a_0^2H_0^2}f^2 \mathcal{P}_0(f), 
\end{equation}
where $a_0 = a(\tau_0) = 1$. We find the power spectrum in MG by finding the form of the enhancement factor that will amplify the GR power spectrum. The power spectrum, in general, is defined as 
\begin{equation} \label{eqn:power_spec}
    \mathcal{P}(\tau,k) = \frac{4k^3 |v_k(\tau)|^2}{\pi^2 M_{\rm Pl}^2 a^2(\tau)}.
\end{equation}
We are interested in $v_k(\tau)$ when it is in the massless regime, for large $\tau$. Solving the equation of motion (\ref{eqn:eom}), applying the appropriate initial and boundary conditions, leads to the following: \begin{equation} \label{eqn:mode_phase_3}
        v_k^{\rm (iii)}(\tau) = \frac{2}{k}\sqrt{\frac{m\tau_m}{\pi H_{\inf} \tau_r^2}}\left[D_1 \cos(k\tau) + D_2 \sin(k\tau) \right],
    \end{equation}
where $D_1 = -\sin(k\tau_m)[C_2\cos(\Lambda + \pi/8) - C_1 \sin(\Lambda -\pi/8)]$, $D_2 = \cos(k\tau_m)[C_2\cos(\Lambda + \pi/8) - C_1 \sin(\Lambda -\pi/8)]$, $C_{1,2} = -i\sqrt{\pi} 2^{-\frac{\tau}{2}+\nu}(k\tau_r)^{-\nu}\Gamma(\nu)[\frac{2m}{H_{\inf}}J_{\mp \frac{3}{4}}(\frac{m}{2H_{\inf}})\pm (1-2\nu)J_{\pm \frac{1}{4}}(\frac{m}{2H_{\inf}}) ] $, and $\Lambda = m\tau_m^2/(2H_{\inf}\tau_r^2)$ \cite{Fujita:2018ehq}. With the assumptions that $m/H_{\rm inf} \approx \mathcal{O}(1)$ and $\Lambda \gg k\tau_m$, we find that the squared amplitude of the mode function in the massless phase is approximately 
\begin{equation}\label{eqn:mode_squared}
    |v_k(\tau)|^2 \sim \frac{\tau_m}{2}(k \tau_r)^{-2\nu - 2}.
\end{equation}
Thus, we find that the power spectrum in MG, in terms of the GR power spectrum, is 
\begin{equation}\label{eqn:p_sfm}
    \mathcal{P}(\tau, k) \sim \frac{\tau_m}{\tau_r}(k\tau_r)^{3-2\nu}\mathcal{P}_{\text{GR}}(\tau,k), 
\end{equation}
where $\mathcal{P}_{\text{GR}}$ is the power spectrum of the massless tensor modes from inflation, and $\nu$ is defined by 
\begin{equation}\label{eqn:nu}
    \nu = \sqrt{\frac{9}{4} - \frac{m^2}{H_{\inf}^2}}.
\end{equation}
This expression for the power spectrum is found by solving the equation of motion for the mode function and finding the expression for $\bar{h}_k$ deep in the massless phase (see Sec.\ III(iii) of Ref.\ \cite{Fujita:2018ehq} for more details). As for the GR power spectrum, we can approximate it by using the transfer function detailed in Ref.\ \cite{Kuroyanagi:2014nba}. To briefly recount, the power spectrum in the GR case is 
\begin{equation}\label{eqn:p_gr_sfm}
    \mathcal{P}_{\text{GR}} = \mathcal{P}^{\text{prim}}_{T} T^2_T(k),
\end{equation}
where $\mathcal{P}^{\text{prim}}_{T}$ is the primordial tensor power spectrum, and $T_{T}$ is the transfer function describing the standard cosmological scenario. $\mathcal{P}^{\text{prim}}_{T}$ is defined in Eq.\ (7) of Ref.\ \cite{Kuroyanagi:2014nba} as follows:
\begin{equation}\label{eqn:pt}
    \mathcal{P}_{T}^{\text{prim}}(k) = A_T(k_{\text{ref}})\left(\frac{k}{k_{\text{ref}}}\right)^{n_T},
\end{equation}
where $A_T(k_{\text{ref}})$ is the amplitude at the reference scale,
\hspace{-1em}measured to be precisely $4.4\times 10^{10}$, and $n_T$ is the spectral index. The reference scale $k_{\text{ref}}$ is chosen to be 0.01 Mpc$\/^{-1}$. $T_{T}$ is defined in Eq.\ (12) of Ref.\ \cite{Kuroyanagi:2014nba} in the following way:
\begin{equation}\label{eqn:tt}
    T_T^2(k) = \Omega_m^2 \frac{g_*(T_\text{in})}{g_{*0}} \frac{g_{*s0}^{4/3}}{g_{*s}^{4/3}(T_{\text{in}})} \frac{9j_1^2(k\tau_0)}{(k\tau_0)^2}T_1^2(x_{\text{eq}}) T_2^2(x_R),
\end{equation}
where $g_{*}(T_\text{in})$ and $g_{*0}$ are the relativistic degrees of freedom at the inflation temperature scale and the present respectively, $g_{*s}(T_\text{in})$ and $g_{*s0}$ are their counterparts for entropy, $j_1(k\tau_0)$ is the first spherical Bessel function whose approximation $j_1(k\tau_0) \simeq 1/(\sqrt{2}k\tau_0)$ is used, the fitting functions are empirically found to be $T_1^2(x) = 1+1.57x+3.42x^2$ and $T_2^2(x) = (1-0.22x^{1.5} + 0.65x^2)^{-1}$, and $x_i \equiv k/k_i$. The values for all of the constants and the forms of the functions are taken from Sec. 2.1 of Ref.\ \cite{Kuroyanagi:2014nba}. 
Since we are interested in the energy density at the present, we consider $\Omega_{\text{GW},0}(f) = \Omega_\text{GW}(\tau_0,f)$,

\begin{figure}
\includegraphics[width=\linewidth]{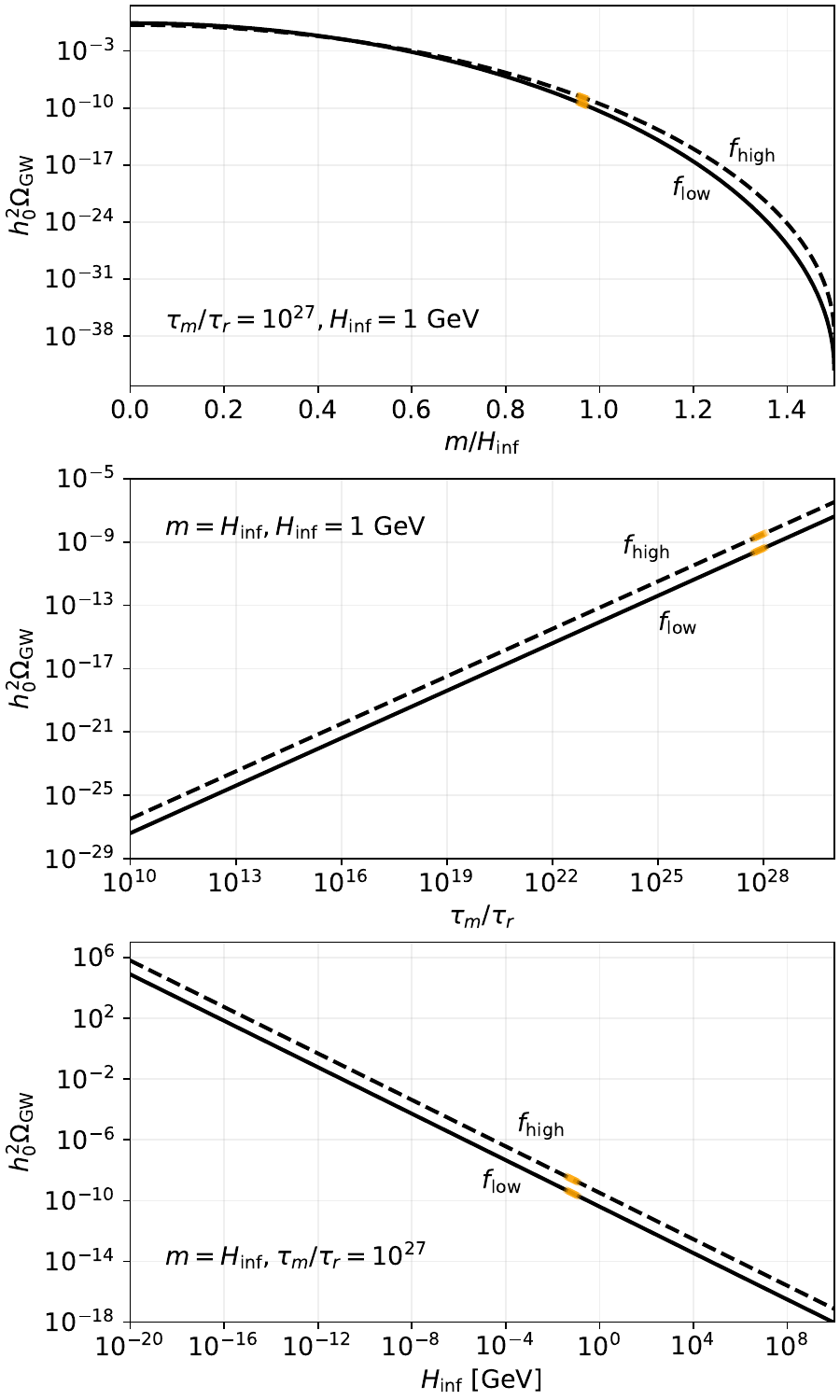}
\caption{We plot $h_0^2\Omega_{\text{GW},0}$ as a function of $m/H_{\inf}$ (top), $\tau_m/\tau_r$ (middle), and $H_{\inf}$ (bottom). The solid lines are the energy densities at the lower bound of NG15, $f_{\rm low} \approx 1.98\times 10^{-9}$ Hz, and the dotted lines are at the lower bound of NG15, $f_{\rm high} \approx 5.93\times 10^{-8}$ Hz. The orange region on each line is the area in the parameter space that produces the observed NG15 energy densities within 1$-$3$\sigma$ of uncertainty.}
\label{fig:contours}
\end{figure}

\begin{equation}\label{eqn:om_gw_0}
    \Omega_{\text{GW},0}(f) = \frac{\pi^2f^2}{3a_0^2 H_0^2}\frac{\tau_m}{\tau_r}(k\tau_r)^{3-2\nu}\mathcal{P}_{\text{GR}}(k).
\end{equation}

In anticipation of our consideration of certain parameters in the next section, we look at the behavior of the energy density as a function of different parameters. Figure \ref{fig:contours} shows how $\Omega_{\text{GW},0}$ varies with changing $m$, $\tau_m$, and $H_{\inf}$. In order to find the region in the parameter space that is able to produce the energy density observed in NG15, we obtain the parameters that fall within 1$-$3 $\sigma$ of the NG15 at the lower and upper bound of frequency. Figure \ref{fig:contours} shows a small region, in orange, in each plot that corresponds to these energy densities for a particular set of parameters. By looking at the energy density dependence on $m, \tau_m$, and $H_{\inf}$, we can find the values of the parameters that reproduce the signal.

\section{Results}\label{sec:results}
We now discuss the region of the parameter space that can potentially explain the signals from NG15. The constraints we can place on the SFM based on NG15 are done by seeing how we can change the parameters $M_{\text{GW}}, \tau_m$, and $H_{\inf}$ to fit the signal. Our initial approach of fixing $H_{\inf}$ to $10^8 \GeV$, like in Ref.\ \cite{Fujita:2018ehq} and letting $m$ and $\tau_m$ vary did not produce any energy density from inflation that could possibly reproduce NG15. We see in the middle plot of Figure \ref{fig:contours} that increasing $\tau_m$ uniformly increases the primordial energy density, at the frequency bounds chosen.

\begin{figure*}
\includegraphics[width=\textwidth]{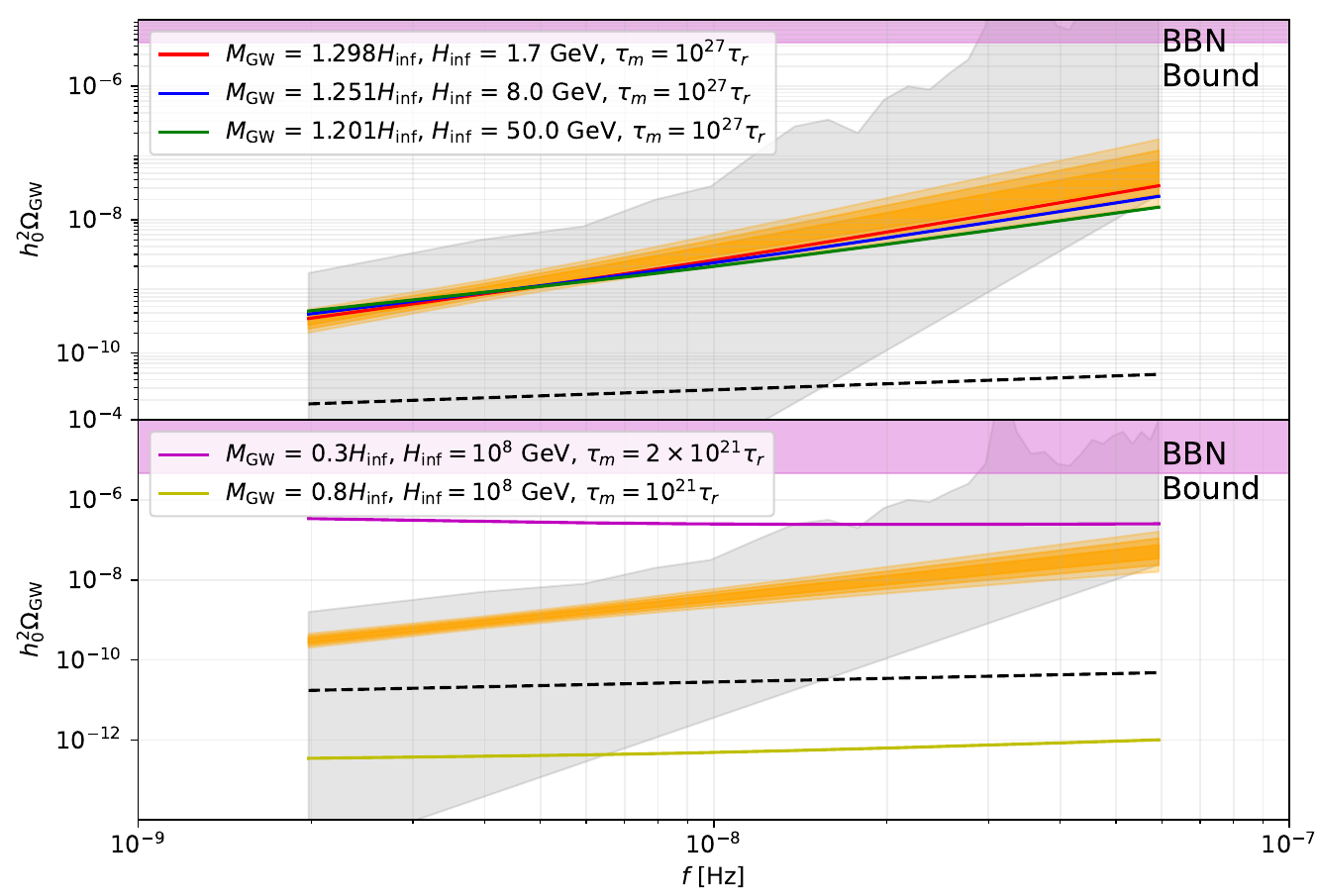} 
\caption{SGWB produced by the SFM model. Both panels: we show the BBN excluded region shaded in purple at the top, the periodogram for a Hellings-Down-correlated free spectral process \cite{Agazie:2023} in shaded gray, the $1\sigma$, $2\sigma$, and $3\sigma$ posterior medians for NG15 \cite{Agazie:2023} in darker to lighter orange, respectively, and the GWB spectrum produced by an astrophysical population of inspiraling SMBHBs with parameters detailed in Eq.\ (A1) of Ref.\ \cite{Afzal:2023} as a black dotted line. Top: The red curve is the GWB spectra fitted to the $1\sigma$ posterior, the blue curve is fitted to the $2\sigma$ posterior, and the green curve is fitted to the $3\sigma$ posterior. Bottom: the purple curve is the energy density that respects the BBN bound for high frequency and passes through the upper limit of the free spectral process of the data, and the golden curve is the energy density that respects the CMB bound for low frequency and passes through the lower limit of the free spectral process of the data.} 
\label{fig:GWB}
\end{figure*}

By varying the other two parameters, $M_{\text{GW}}$ and $H_{\inf}$, we found resulting energy densities that lie within $1\sigma$, $2\sigma$, and $3\sigma$ of the power law posterior of the signal. Our values for the parameters are $M_{\text{GW}} = 1.298H_{\inf}$ and $H_{\inf} =  1.7 \GeV$ to stay within $1\sigma$ of the posterior, $M_{\text{GW}} = 1.251H_{\inf}$ and $H_{\inf} = 8.0 \GeV$ to stay within $2\sigma$ of the posterior, and $M_{\text{GW}} = 1.201H_{\inf}$ and $H_{\inf} = 50.0 \GeV$ to stay within $3\sigma$ of the posterior. We have kept the ratio $\tau_m/\tau_r$ constant for these three energy densities, although the magnitude of $\tau_m$ changes due to the change in $\tau_r$. 

The bound on $\Omega_{GW,0}$ made from the big bang nucleosynthesis (BBN) limits comes from the explicit limit on the integral of the energy density [Eq.\ (3.3) of Ref.\ \cite{Tanin:2020qjw} ], 
\begin{equation} \label{eqn:bbn_int}
    \int_{f = f_{BBN}}^{f = \infty}\mbox{d}(\ln{f}) h_0^2\Omega_{GW,0}(f) \ \lesssim\ 1\times 10^{-6},
\end{equation} 
where $f_{BBN}$ is the present-day frequency at which the GWs at the time of BBN were inside the horizon and therefore contribute to the expansion of the Universe and is approximately $1.5\times 10^{-11}$ Hz [Eq.\ (22.290) of Ref.\ \cite{Maggiore:v2} ]. We can reformulate Eq.\ \eq{eqn:bbn_int} as a bound on the energy density rather than the integral of the energy density, since in most cosmological mechanisms, including the one considered in this paper, the spectrum covers at least one decade of frequency. So thus we have the bound,
\begin{equation}
    \Omega_{\text{GW}}(f) \lesssim 1 \times 10^{-6},\ f > f_{\text{BBN}}.
\end{equation}

\begin{figure}[ht]
    \includegraphics[width=\linewidth]{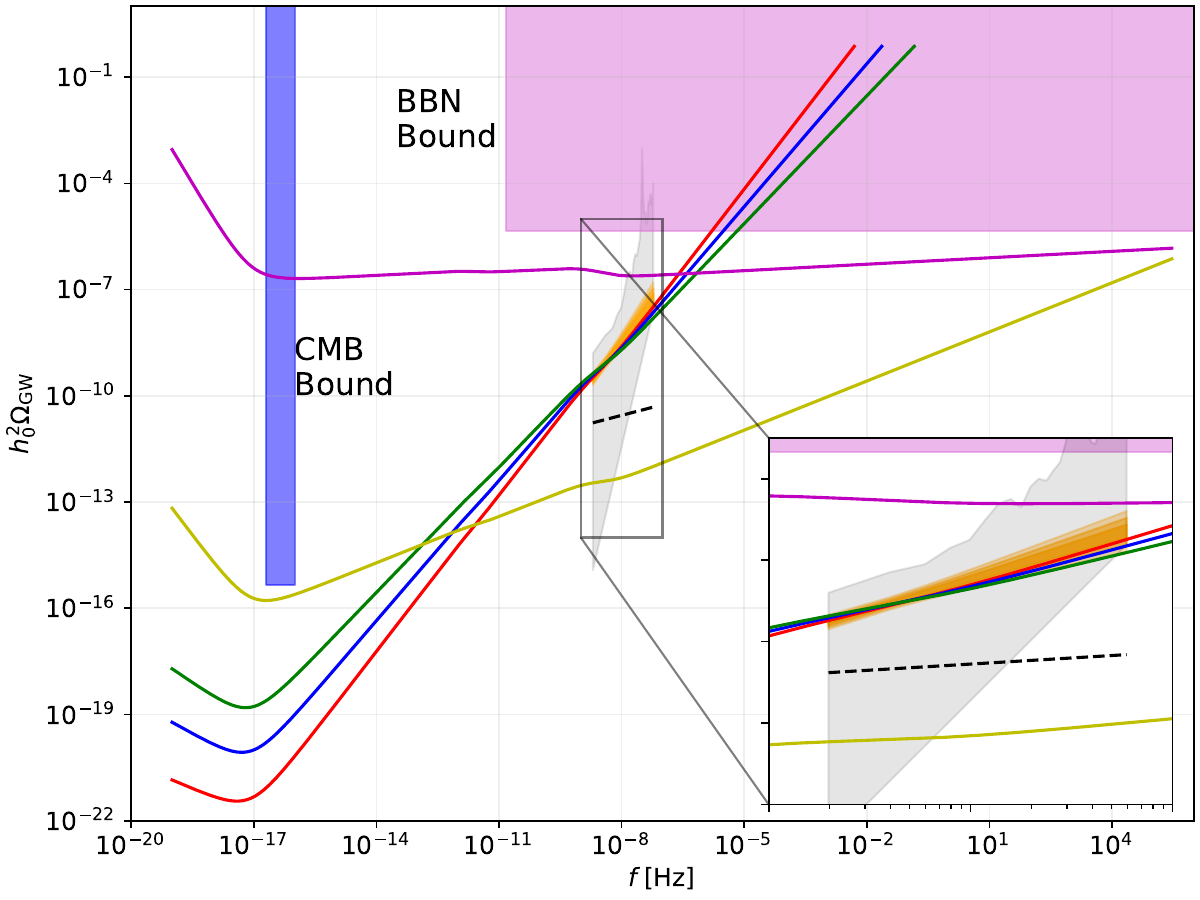}
    \caption{The energy densities in Fig.\ \ref{fig:GWB} plotted over the frequency range from the scale corresponding to matter-radiation inequality ($\sim 3\times10^{-16}$ Hz) to the inflationary UV cutoff ($\sim 1/(2\pi\tau_r)$ \cite{Fujita:2018ehq}). The red, blue, green, purple, and golden curves correspond to the energy densities of the same colors in Fig.\ \ref{fig:GWB}. The blue-shaded region on the left is the excluded region from the CMB bound. The BBN bound region may be ignored if the graviton remains massive during BBN.} 
    \label{fig:supp}
\end{figure}

Although it may appear that the energy densities we consider in Figure \ref{fig:GWB} violate the BBN bound for higher frequencies since the energy densities in the full range of frequencies pass into the forbidden energy range as we see in Figure \ref{fig:supp}, we keep in mind that the BBN bound only applies to GWs that were blue-tilted by massive gravitons before BBN took place. If the graviton were massive during BBN, 
\begin{equation}\label{eqn:massive_bbn}
    \frac{\tau_m}{\tau_r} \gtrsim \frac{\tau_{\text{BBN}}}{\tau_r} = \sqrt{\frac{H_{\inf}}{H_{\text{BBN}}}},
\end{equation}
then we should be able to relax the BBN bound. This is because gravitons do not contribute to the relativistic degrees of freedom during BBN \cite{Fujita:2018ehq}, since they would have had a significant mass, meaning that they contribute to nonrelativistic species.

We note that if we still wish to conservatively respect the BBN bound in the same way as Ref.\ \cite{Fujita:2018ehq} and still have our theory of MG be consistent with NG15, we can use the parameters investigated by Ref.\ \cite{Fujita:2018ehq} with some slight modifications. We see that in order to partially produce the signal in the higher energy density and frequency regime, as shown by the purple curve in Figs.\ \ref{fig:GWB} and \ref{fig:supp}, the cosmic microwave background (CMB) bound must be violated. On the other hand, we find that while attempting to partially produce the signal in the lower energy density and frequency regime, as shown by the golden curve in Figs.\ \ref{fig:GWB} and \ref{fig:supp}, the SMBHB power spectrum turns out to be a better fit to the signal. The requirement to respect both bounds to reproduce the signal seems to be mutually exclusive. But if we do not conservatively respect the BBN bound, then it appears that we can achieve good agreement with the signal, with the caveat being that the cutoff time for the graviton mass is deep into the matter-dominated era.

\section{Discussion}\label{sec:discussion}
In this paper, we discussed how the step function mass model of MG can reproduce NG15. We note that the selection of the parameters in the step function mass model leads to energy densities that violate the BBN bound for higher frequencies, specifically in the region of frequencies approximately greater than $10^{-6}$ Hz. We propose that there could be some mechanism that suppresses the energy density of GWs from the primordial era for those frequencies. A mechanism analogous to the damping of the energy density from the free-streaming neutrinos \cite{Durrer:1997ta,Weinberg:2003ur} could be behind such a suppression necessary to obey the BBN bound, if this model of MG is to be the dominant source of the background. For example, the propagation of GWs in the turbulent primordial radiation-dominated plasma\footnote{Turbulent sources induce GWs through their anisotropic stress tensor, with $\Omega_{\rm GW} \propto f^{-8/3}$ for high frequencies.} \cite{RoperPol:2019wvy} may provide the necessary suppression at the short wavelength regime \cite{Deryagin:1986}. 

The values for $H_{\inf}$ we find that reproduce the SGWB are very low. Various models of inflation with sufficiently low $H_{\inf}$ have been proposed \cite{Nakayama:2011ri, Takahashi:2018tdu, Drees:2021wgd}. In models of inflation such as these, a significant blue-tilt is required to explain the SGWB, presenting a challenge to purely inflationary explanations \cite{Vagnozzi:2023lwo}. Additionally, the existence of the SGWB would most likely rule out standard inflationary models due to their dilution of the SGWB to an undetectable level \cite{Vagnozzi:2022qmc}. Thus, we can look to nonstandard inflationary scenarios that include features like the breaking of the slow-roll consistency relation leading to the desired blue-tilted spectrum \cite{EPTA:2023xxk}, such as MTMG.

Observations of GWs have placed progressively tighter constraints on the possibility of gravitons with a very low but constant nonzero mass \cite{Bernardo:2023mxc, Wu:2023rib, Wang:2023div, deRham:2016nuf}, including much lower bounds with time-independent MTMG \cite{DeFelice:2021trp,DeFelice:2023bwq}. In addition, these models of MG do not provide a sufficiently blue-tilted spectrum to explain the SGWB. Therefore, the more plausible formulation of MG is the scenario in which the graviton mass is a function of time and the graviton used to have a significant mass, in order to provide sufficient blue-tilt. We have only considered a step function as the time-dependent function, but more complicated functions are possible. The mechanism behind such a mass decay would come from the exact nature of the phase transition of the gravitational field. It may be fruitful to pursue an investigation to place constraints on the specific evolution of the mass during the phase transition. Then, we would be able to probe the mass evolution and shed insight into the time-dependent behavior beyond a step function.

Additionally, further observations that place constraints on the Hubble rate of expansion, the scale factor, and the time associated with inflation would be able to constrain the parameters of this theory. In addition to signals we have already observed with interferometers, we expect a more drastic suppression for higher frequencies than we discuss. Future work may investigate the nature of this suppression and propose plausible mechanisms for it.

\vspace{3mm}
The NANOGrav 15-year data used in this paper are available at NANOGrav \cite{NANOGrav:2023}. Source code to reproduce all of the figures in this paper is available in our GitHub repository \cite{Choi:2024}. 

\begin{center}
    \textbf{ACKNOWLEDGMENTS}    
\end{center}
\ \ \ We thank Sachiko Kuroyanagi and Shinji Mukohyama for helpful discussions related to the theoretical foundation of SFM, \cite{Fujita:2018ehq}, their useful comments, and their feedback. We appreciate discussions from Axel Brandenburg, Neil J.\ Cornish, and Arthur B.\ Kosowsky. We thank Emma Clarke, Jeffrey S.\ Hazboun, and William G.\ Lamb for their help with plotting NG15, and Sayan Mandal for interpreting our results. We thank the organizers and participants of the PITT PACC workshop ``Unravelling the Universe with Pulsar Timing Arrays Workshop" during which part of this paper has been written. T.K.\ and M.G.\ acknowledge partial support from the NASA Astrophysics Theory Program (ATP) Award No.\ 80NSSC22K0825. T.K.\ acknowledges partial support from NSF Astronomy and Astrophysics Research Grants (AAG) Award No.\ 2307698.

\bibliographystyle{apsrev4-2_edited}
\bibliography{refs}

\clearpage
\end{document}